# Machine learning shadowgraph for particle size and shape characterization

Jiaqi Li[1,2], Siyao Shao[1,2], Jiarong Hong[1,2, *]

1. Saint Anthony Falls Laboratory, 2 3rd Ave SE, University of Minnesota, Minneapolis, MN, 55414, USA
2. Department of Mechanical Engineering, University of Minnesota, Minneapolis, MN, 55414, USA

* Email address of the corresponding author: jhong@umn.edu

## Abstract

Conventional image processing for particle shadow image is usually time-consuming and suffers degraded image segmentation when dealing with the images consisting of complex-shaped and clustered particles with varying backgrounds. In this paper, we introduce a robust learning-based method using a single convolution neural network (CNN) for analyzing particle shadow images. Our approach employs a two-channel-output U-net model to generate a binary particle image and a particle centroid image. The binary particle image is subsequently segmented through marker-controlled watershed approach with particle centroid image as the marker image. The assessment of this method on both synthetic and experimental bubble images has shown better performance compared to the state-of-art non-machine-learning method. The proposed machine learning shadow image processing approach provides a promising tool for real-time particle image analysis.

## 1. Introduction

Characterizing size and shape distribution of particles (e.g., bubbles, droplets, cells, sediments etc.) is critical in a broad range of applications such as, hydrocarbon synthesis (Maretto and Krishna 1999) and spray-assisted internal combustion engines (Hayashi et al. 2011), drift control of pesticide spray (Gil et al. 2015 and Kumar et al. 2019), particle entrainment over sediment surfaces (McEwan et al. 2000), hydroturbine aeration (Karn et al. 2015a), and blood disease detection (Di Rubeto et al. 2000). Shadowgraphy is a popular imaging-based technique for extracting particle size distribution and their velocities (Estevadeordal and Goss 2005). It captures shadow images of particles generated from backlighting with a camera and uses image processing to extract the size and motion of particles.

Particle image segmentation is the most challenging step in the image processing of shadowgraphy (Karn et al. 2015b). A number of segmentation methods have been proposed based on different metrics, but in general, conventional algorithms usually suffer degraded performance with complex-shaped particles and clusters and varying background noises due to different image settings and particle characteristics. Moreover, these conventional approaches usually require substantial tuning of segmentation parameters (e.g., the threshold value for image binarization, morphological operator parameters for segmentation) and sophisticated processing steps which are time consuming. Specifically, shape-criterion-based methods, such as elliptical fitting (Honkanen et al. 2005) and concentric circular arrangements method (Strokina et al 2012), rely on assumptions of particle shape and size and have limited performance for particles with complex geometry. A number of algorithms demarcate neighboring particles by detecting concave points along the perimeter of the particle clusters (Bai et al. 2009 and Zhong et al. 2016). Such method is not only computationally intensive but susceptible to image noise which deteriorates concave point detection. While watershed algorithm employed in Lau et al. (2013) could yield more robust particle segmentation, it still suffers oversegmentation issue when the boundary and internal intensity of particles varies substantially, especially for the cases of irregular shape particles. Recently, Karn et al. (2015b) integrated multiple segmentation approaches, including extended H-minima transform for image binarization, morphological operators for in-focus particle clusters as well as watershed algorithm for out-of-focus clusters to achieve optimal segmentation results. However, their method still requires threshold tuning to cope with varying background noise associated with heterogeneous distribution of bubbles in space, and cannot accurately capture the shape of irregular bubbles due to the inherent elliptical assumption employed in their method.

Machine learning has become a prevailing technique for visual recognition and image segmentation owing to the recent development of deep convolutional neural networks (CNN) (Ciresan et al. 2012 and Shen et al. 2017). For shadowgraphy, a number of studies have applied machine learning to the detection and analysis of particles (Ilonen et al. 2018, Poletaev et al. 2020, and Haas et al. 2020). Both Ilonen et al. (2018) and Poletaev et al. (2020) employ CNN trained on an image of single bubble, and use the trained model to slide over the entire bubble image to extract each individual bubble at the center of the sliding window. Specifically, Ilonen et al. (2018) has shown that the CNN-based method yields higher accuracy than conventional approaches (e.g., CCA). However, their method requires manually-labeled data for model training and testing, which is labor-intensive and can involve human errors even for a small number of data present in their paper. More importantly, their learning-based method yields only an 82.2% particle extraction rate which is still insufficient for many practical applications. Poletaev et al. (2020) combined three consecutive CNNs, including a bubble detector CNN to identify the particle regions, a center finder to extract particle centroids afterwards and a denoising autoencoder multi-layer perceptron network to filter particles from background noise in the end. Although their method can achieve a higher overall particle extraction rate of 92.2% in comparison to Ilonen et al. (2018), it has a slow processing time due to the usage of multiple CNNs with sliding windows which only extracts one bubble at a time. To address this issue, Haas et al (2020) applied a fast region-based CNN (RCNN) to locate the bubble centroids and a shape regression CNN to measure bubble shape. Note that a quarter of the training set used in this study is from a synthetic bubble dataset generated through a learning-based approach which can produce more realistic synthetic images compared to conventional approaches (Fu et al. 2019). However, their method still requires training of three CNNs which is time consuming and has not been fully validated through a systematic comparison with ground truth with quantitative results (i.e., extraction rate and accuracy have been assessed).

To address abovementioned limitations, in the present study, we will introduce a U-net based single CNN to achieve fast and accurate analysis of particle images from shadowgraph. The architecture of this learning-based methodology was first developed by Shao et al. (2020) for tracer field extraction based on digital holography and has been applied to holographic particle sizing tasks (Shao et al. 2019). The present paper is structured as follows: Section 2 provides a description of the proposed learning-based shadow image analysis approach. Successively, an assessment of the proposed method is conducted using both synthetic and experimental bubble images in Section 3 (including a comparison with the non-machine-learning method introduced in Karn et al. (2015b)) with a summary and discussion provided in Section 4.

## 2. Methodology

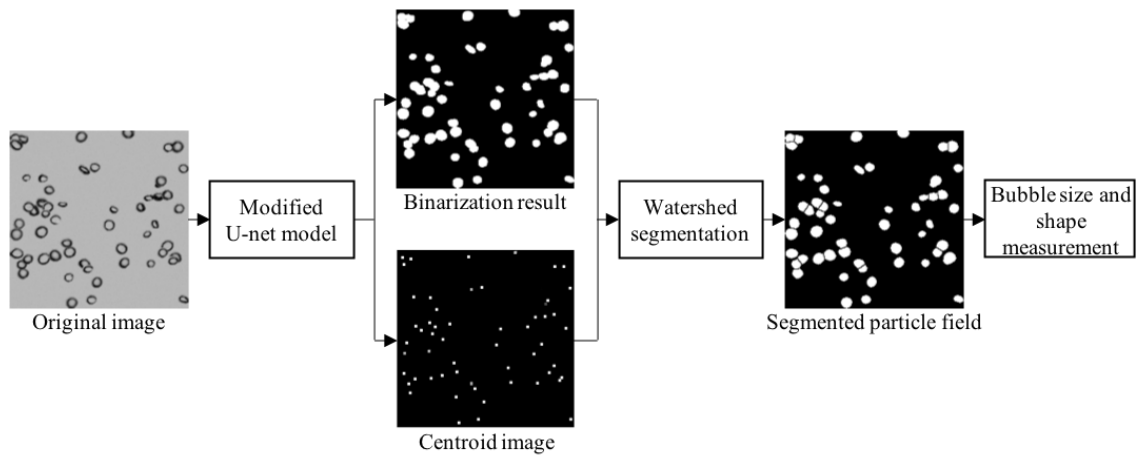

Figure 1. Outline of the proposed method for particle shadow image analysis.

Fig. 1 presents the general methodology of the proposed learning-based shadow image processing algorithms. The proposed method consists of three steps: (i) shadow image binarization and particle centroids detection using a two-channel-output CNN; (ii) segmentation of particle binary image through marker-controlled watershed algorithm using centroid image as marker image; (iii) region properties analysis for particle size and shape measurements.

The machine learning model used in the present approach is a modified U-net developed by Shao et al. (2020) (Fig. 2). The U-net has been originally proposed for biomedical image segmentation (Ronneberger et al. 2015). Compared to sliding window CNN (Ciresan et al. 2012) and densely-connected CNN (Dolz et al. 2018), skip connection used in U-net reduces the complexity of the network to produce an equally or even better image segmentation with robust training and simple post-processing. To further improve training speed and measurement accuracy, the residual connection has been incorporated into the original U-net in our design. Additionally, commonly-used ReLU (rectified linear unit) activation function is replaced with Swish (Sigmoid-weighted linear unit) except the last layer, which improves the performance of the machine learning model by increasing the number of trainable parameters within the model (Ramachandran et al. 2017). The input of the model is the shadow bubble image and the two output channels are a binarized bubble image and an image contains all bubble centroid regions. It is worth noting that the last convolutional layer of the U-net model uses a ReLU activation which provides a sparse and clean output of both particle binary image channel and particle centroid channel. Such feature greatly reduces the noise in the model output especially for the cases with non-uniform background occurring in the shadowgraph of large field of view.

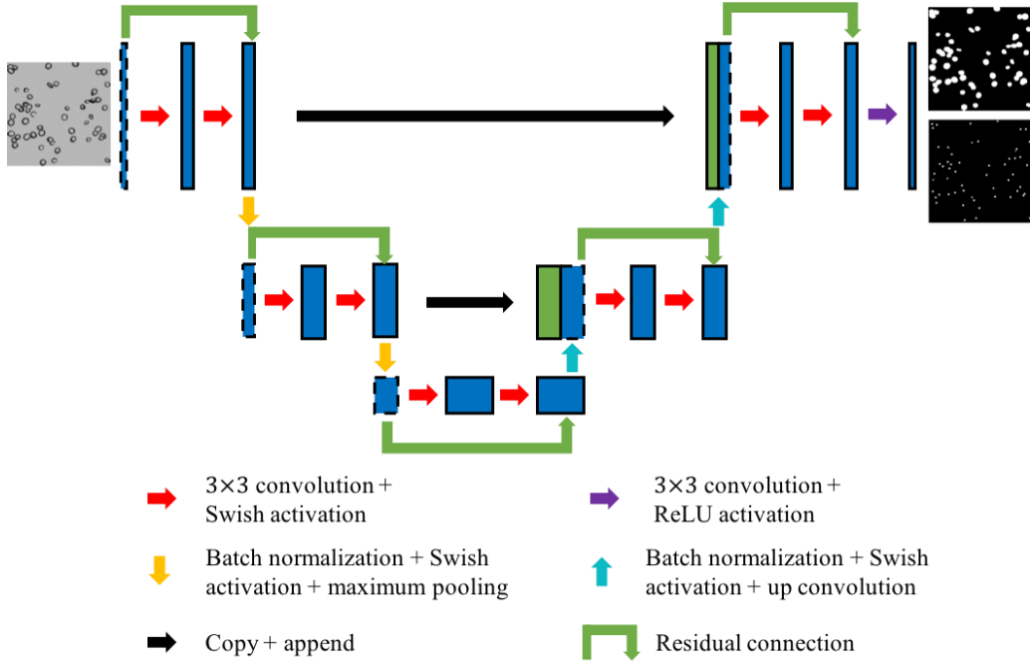

Figure 2. U-net architecture applied in the shadow image analysis.

Unlike previous particle sizing task using holography (Shao et al. 2019), a binary cross entropy loss function is selected in the training of the binary particle image channel (Eq. 1). Binary cross entropy loss is usually used in the object classification with binary outputs since it can converge fast which greatly reduces the training time (Ronneberger et al. 2015 and Janocha and Czarnecki 2017). In the equation, $X_{i,j}$ is ground truth and $Y_{i,j}$ is model prediction. In the proposed method, the machine learning model classifies the pixel into a background or a particle pixel in the binary particle image channel.

$$L = -\frac{1}{2N}\sum_{i=1}^{N}\sum_{j=1}^{N}(X_{i,j} \times log(Y_{i,j}) + (1-X_{i,j}) \times log(1-Y_{i,j})) \qquad (1)$$

A total variation (TV) regularized mean square error (MSE) loss (Eq. 2) is employed in the centroid channel. As shown in Eq. 3, TV is the sum of first order gradients over the image of size $Nx$ by $Ny$. TV regularized MSE loss has been previously adopted for 2D particle centroid channels for machine learning based holography analysis (Shao et al. 2019 and 2020). It helps remove false detected particle centroids and provides a clean and smooth background by minimizing the first order gradient of the output images. In our case, the parameter $\alpha$ controlling the smoothness of model output is set as 0.0001.

$$L = (1-\alpha) \times \|Y-X\|^2 + \alpha \times [TV(Y)]^2 \qquad (2)$$

$$TV(Y) = \sum_{i=1}^{Nx}\sum_{j=1}^{Ny}\sqrt{(Y_{i,j}-Y_{i-1,j})^2 + (Y_{i,j}-Y_{i,j-1})^2} \qquad (3)$$

The binary bubble image obtained from U-net is subsequently segmented by a marker-controlled watershed algorithm (Gonzalez et al. 2009). The binary channel output of particles is segmented through watershed with distance transform using the particle centroid channel as marker image. The method only employs a perpendicular bisector of the line connecting two adjacent marker centroids as the segmentation line, which largely eliminates the oversegmentation. The performance of such segmentation will be discussed in the following section in detail.

Measurement of bubble size and shape follows the approach provided by Karn et al. (2015b). Bubble size is defined by area equivalent radius $R = \sqrt{A/\pi}$, where $A$ is the area of each segmented bubble. The particle shape is quantified by the aspect ratio calculated as the ratio of major and minor axes lengths of the bubbles.

## 3. Assessment of proposed method

The machine learning model is trained using synthetic bubble images from BubGAN database with corresponding labels (Fu et al. 2019). This method uses generative adversarial network (GAN) (Radford et al. 2015) to train a model to generate bubble shadow images. BubGAN training takes aspect ratio, rotation angle, circularity, edge ratio and diameter of individual bubbles as the input with additional real bubble images from a bubble column as reference images. During the training, a generator model is used to generate synthetic bubble image from input shape parameters of individual bubbles and a discriminator model classifies whether the synthetic bubble images from the generator belong to the real bubble images from reference dataset. By minimizing the visual difference between synthetic and real bubbles images, the discriminator eventually fails to distinguish synthetic bubble images from the real ones and the generated single bubble images are subsequently stitched to bubble field images based on user-defined bubble size, shape, and location distributions (an example shown as Fig. 3) (Fu et al. 2019). As mentioned in the introduction, BubGAN generates more realistic synthetic bubble images for machine learning training (Hass et al. 2020) compared to conventional method by applying modelled pixel intensity variation on simple shapes such as circles and ellipses (Ilonen et al. 2014 and Karn et al. 2015b). The machine learning model used for bubble image analysis in this section is trained on a 2000 synthetic images and their corresponding labels. Each synthetic image is 256×256-pixel in size and consists an average of 100 bubbles with equivalent radius ranging from 4 to 8 pixels (average 6 pixels) and void fraction of 0.07. The model is trained for 140 epochs using TensorFlow 2.0 on an Nvidia RTX 2070 GPU. Adam optimizer is applied in the training with default settings (Kingma and Ba 2014). Training process takes around 7 hours. After training, the model is tested both on synthetic bubble images and experimental data using a single CPU processor.

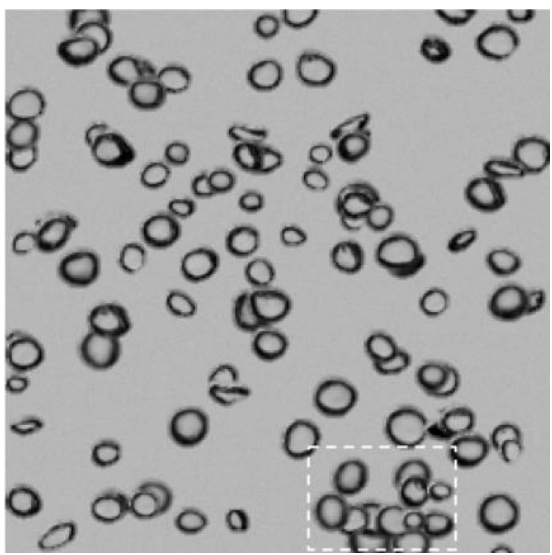

Figure 3. Example of synthetic bubble image. The dashed rectangle box marks an example of bubble cluster.

## 3.1 Assessment through synthetic data

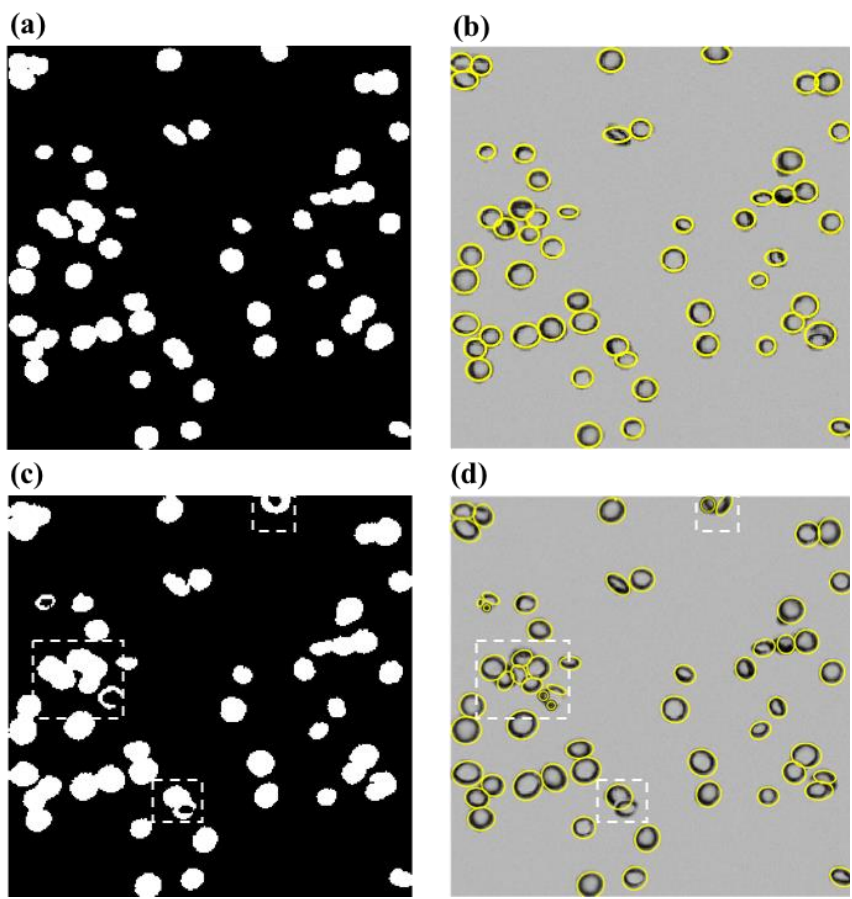

Figure 4. Sample of binarization output (a) and segmentation result with yellow line labelling bubble edges (b) from the proposed method, as well as binarization output (c) and segmentation result (d) from the method proposed by Karn et al. (2015b). Note that the dashed rectangles mark the inaccurate bubble cluster segmentations (d) due to binarization errors (c).

The proposed method is first assessed using 200 BubGAN synthetic images. In the testing dataset, each synthetic image has an average of 50 bubbles with equivalent radius ranging from 3 to 9 pixels (average 7 pixels) and void fraction of 0.055. Fig. 4 shows sample outputs of bubble binarization and segmentation results from the proposed method and Karn et al. (2015b) approach. The proposed method exhibits nearly perfect capture of all the bubbles and segmentation of clusters (Fig. 4a and b). In comparison, the conventional methods such as Karn et al. (2015b) show some drawbacks associated with the dependence of their performance on bubble shape and texture (i.e. grayscale intensity within the bubble due to varying light conditions and bubble orientation). Specifically, for some bubbles, parts of the bubble edges are absent from the binarized bubble image using Karn et al. method and the bubble edges in the clusters are not well matched with the ground truth (Fig. 4c). This binarization error is primarily due to the sensitive performance of binarization to the image intensity variation near the edge of the bubbles. Note that the threshold used in Karn et al. method has been already optimized according to the image intensity variation locally and temporally. Subsequently, some of these incorrect binarized bubbles and clusters with irregular shapes are oversegmented since non-machine-learning approach assume that bubbles are either spherical or elliptical. Comparatively, the binarization of image through machine learning does not take any predefined image threshold which shows noticeable improvement of the image binarization. Additionally, the marker-controlled watershed algorithm could avoid oversegmentation by marking bubble centroid regions using bubble centroid output produced from the learning model without assumptions.

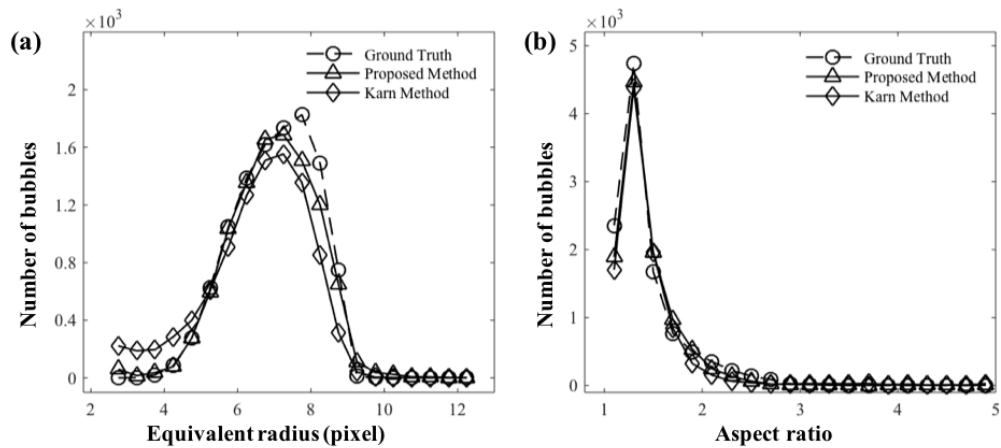

Figure 5. (a) Comparison of ground truth and measured size distribution from proposed method and Karn et al. (2015b) method and (b) comparison of aspect ratio.

Fig. 5 shows a comparison of size distribution and aspect ratio distribution among ground truth, results from learning-based method and conventional approach (Karn et al. 2015b). The proposed method yields an overall particle extraction rate of 94.7% with only 0.6% false positives in comparison to the 87.3% detection rate with 2.0% false positives of the conventional method. Note that these false positives are primarily belonging to the tiny objects from cluster oversegmentation with equivalent radius less than 3 pixels (Fig.5a) (i.e., minimum bubble size of the ground truth). As shown in Fig. 5a, the bubble size measurement from the proposed method has almost perfect matching with the ground truth in the range from 3 to 7 pixels and the conventional method does not resemble the plot of ground truth in this range due to oversegmentation. Note that in the range of equivalent bubble radius from 7 to 9 pixels, the proposed method underestimates the number of bubbles which can be attributed to overlapping of bubbles in the clusters leading to information loss of bubbles. Further, the proposed approach has shown bubbles with dimension larger than the maximum radius of the ground truth (>9 pixels) which is associated with the undersegmentation of the clusters when the centroids of adjacent bubbles are too closely located to be separated. Nevertheless, our method still provides a reasonable estimation of particle centroid distribution in a cluster which largely avoids the segmentation errors. As shown in Fig. 5b, both the conventional approach and the proposed machine learning method show good agreement of the bubble shape measurement in comparison to the ground truth, and the learning-based method exhibits an improved result. Note that the mismatching of the measured results to the ground truth is due to overlapping of bubbles in clusters which alters the shape of individual bubbles.

### 3.2 Assessment through experiment data

The model trained from synthetic images is subsequently assessed using experimental images of bubbly flow generated by a ventilated hydrofoil. A NACA0015 hydrofoil with 0° angle of attack was installed in a high-speed water tunnel located at Saint Anthony Falls Laboratory (SAFL) of the University of Minnesota. Air ventilation rate is set at 0.5 SLPM (standard liter per minute) and a high-speed camera is set at 377 mm downstream of the hydrofoil for capturing bubble images (Karn et al. 2015b). Fig. 6 shows a sample image from the experiments and the corresponding results from our method and that from Karn et al. (2015b) for comparison. Note that for the implementation of machine learning method, the experimental images are first preprocessed to match the pixel intensity distribution of BubGAN dataset. As shown in Fig. 6a, compared with the synthetic bubble image from BubGAN (Fig. 4), the bubble image from our experiments exhibits different textures and configurations of clusters. Nevertheless, the proposed method, though trained on the synthetic image, captures nearly all isolated bubbles and the majority of bubbles within the clusters (Fig. 6b). In comparison, the conventional method (Karn et al. 2015b) has oversegmention of some isolated bubbles and shows mismatching of bubble edges for large bubbles and clusters (Fig. 6c).

Both methods show some limitations in extracting highly irregular bubbles within the bubble cluster, even though such bubbles can be easily segmented manually. Such limitation can be potentially addressed with an extension of our learning-based method. It is worth noting that the synthetic training dataset is generated using the reference images from bubble column experiments, which is different from our assessments. More important, Fu et al. (2019) used individual bubble shape parameters as the model input which fails to capture the background non-uniformity commonly occurs in large field of view shadowgraph and variation of image characteristics which are closely associated with the void fraction variation in dense bubbly flow. As a result, an intensity matching has to be conducted before processing bubble images using learning-based method. Such preprocessing causes undetected small bubbles due to saturation of pixel intensity. We suggest that for the future applications of our proposed method, GAN model for producing synthetic bubble images should be trained by matching images from conventional synthesis with reference experimental bubble field images. These reference images should be categorized by different experimental facilities and recording settings instead of simply using single bubble images from one facility and limited camera settings. As such, the machine learning models for particle shadowgraph analysis of different applications could have best performance through training on synthetic images best matching real data.

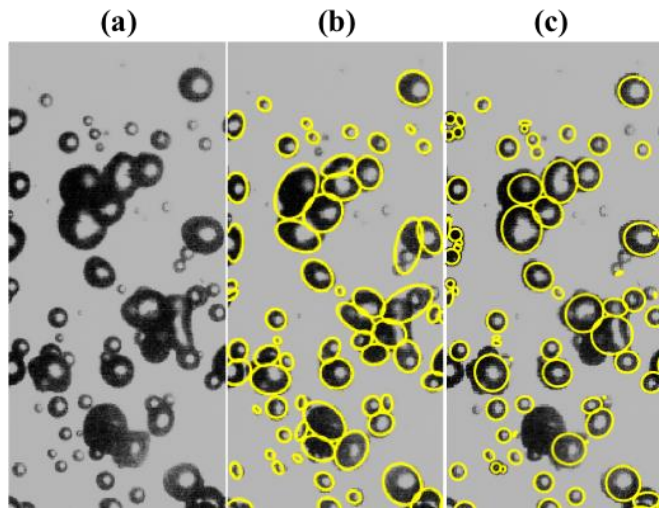

Figure 6. (a)Sample of experiment data after preprocessing; (b)Segmentation result from the proposed method and (c) conventional approach (Karn et al. 2015b). The yellow lines label the edge of detected bubbles.

## 4. Discussion and Summary

In this paper, we proposed a robust particle characterization method based on shadowgraphy using machine learning. We employ a modified U-net architecture as the machine learning model to binarize and locate particles from particle shadow image. With original images as model input, binarized images and particle centroid images can be obtained from a two-channel output. A marker-controlled watershed algorithm has been applied to segment the binarized particle output with centroid image providing markers. A binary cross entropy and a total variation regularized mean square error loss are employed to binary particle output and centroid output during training, respectively. The proposed approach has been assessed through both synthetic and experiment data. The results of synthetic image assessment have demonstrated that the proposed method outperforms the state-of-art conventional shadow image analysis (Karn et al. 2015b) in terms of detection rate and accuracy of particle size measurement. Additionally, the proposed learning-based approach requires minimal parameter tuning to provide optimal analysis of bubble images with more than one order of magnitude increase in processing speed. The assessment of the machine learning method on bubbly flow images generated by a ventilation hydrofoil has shown the potential of directly using machine learning model trained on synthetic data to process real images.

In conclusion, we have developed a novel shadow image characterization method based on a modified U-net architecture with improvement on particle detection rate, size and shape measurement. This method can be applied to a number of online real-time particle image monitoring tasks in industrial and environmental science settings with further optimization of processing pipeline using GPU. Nevertheless, present method has some limitations. First, training of the machine learning model relies on high-quality synthesized particle images which have minimal visual differences with experimental data. This issue can be tackled using learning-based generative adversarial networks (GAN) method to produce high-fidelity synthetic data based on particle images from different facilities captured by various imaging settings. Second, the loss of particle information due to overlapped particles in clusters thwarted further boosting of measurement accuracy through proposed framework. We suggest that through a machine learning prediction approach, such missing information of particles from clusters can be made up to provide more accurate measurements of the particle size and shape.

## Acknowledgements

This work is supported by the Office of Naval Research (Program Manager, Dr. Deborah Nalchajian) under grant No. N000141612755 and the Mechanical Engineering Department Fellowship. The training dataset is from https://github.com/ycfu/BubGAN.